\documentclass[twocolumn,amsmath,amssymb,floatfix,prl,showpacs,footinbib]{revtex4}
\usepackage{amsmath,amssymb,natbib,bm,graphicx,url,psfrag,epsfig}
\usepackage[ansinew]{inputenc}

\newcommand{\be}{\begin{equation}}
\newcommand{\ee}{\end{equation}}
\newcommand{\bes}{\begin{equation}\begin{split}}
\newcommand{\ees}{\end{split}\end{equation}}

\newcommand{\vc}[1]{\mathbf{#1}}

\newcommand{\abs}[1]{\left|#1\right|}

\newcommand{\ket}[1]{\left|\, #1 \, \right\rangle}

\DeclareMathOperator{\intg}{int}

\begin{document}
\title{Franck-Condon blockade and giant Fano factors in transport through single molecules}
\author{Jens Koch}
\author{Felix \surname{von Oppen}}
\affiliation{Institut f\"ur Theoretische Physik, Freie Universit\"at Berlin, Arnimallee 14, 14195 Berlin, Germany}
\date{April 11, 2005}
\begin{abstract}
We show that Franck-Condon physics leads to a significant current suppression at low bias voltages (termed \emph{Franck-Condon blockade}) in transport through single molecules with strong coupling between electronic and vibrational degrees of freedom. Transport in this regime is characterized by remarkably large Fano factors ($10^2$--$10^3$ for realistic parameters), which arise due to avalanche-like transport of electrons. 
Avalanches  occur in a self-similar manner over a wide range of time scales, leading to power-law dependences of the current noise on frequency and vibrational relaxation rate.
\end{abstract}
\pacs{73.23.Hk, 72.70.+m, 73.63.-b, 81.07.Nb}
\maketitle
\emph{Introduction.}---The Fano factor $F$, defined as the ratio $S(\omega=0)/2e\abs{I}$ of the zero-frequency current noise and the classical Schottky result  for Poissonian shot noise \cite{schottky}, contains information about the charge of the current-carrying particles, and about quantum correlations between them. Typically, the Pauli principle leads to  suppressed Fano factors ($F<1$) for fermionic carriers \cite{blanter}. However, super-Poissonian noise has been discovered in systems with instabilities \cite{blanter,kuznetsov, flindt}, and, very recently, with dynamical spin blockade \cite{belzig}. In the present paper, we report surprisingly large Fano factors in transport through single molecules, which originate from the coupling of electronic and vibrational degrees of freedom, but are unrelated to instabilities.

Recently, single-molecule devices with vibrational degrees of freedom (phonons) have received much attention both experimentally \cite{park} and theoretically \cite{schoeller3,aleiner,koch}. Here, we investigate the regime of strong electron-phonon coupling by current-voltage ($IV$) and noise calculations. We find that strong electron-phonon coupling leads to a significant current suppression at low bias voltages, which we term \emph{Franck-Condon blockade}. The striking transport properties of this regime are further elucidated by computing the current fluctuations. We find that systems with weak vibrational relaxation can exhibit \emph{giant Fano factors} of the order of $10^2$--$10^3$. This occurrence of giant Fano factors is due to avalanche-like transport of electrons interrupted by long times without charge transfer. These avalanches occur in a self-similar manner over a wide range of time scales, which is reflected in power-law behavior of the noise amplitude over many orders of magnitude in frequency and relaxation rate. 

\emph{Model.}---Our results are obtained for a model of a molecule weakly coupled to metallic source and drain electrodes. As usual, we take the relaxation in the leads to be sufficiently fast so that their electron distributions have the form of Fermi functions. Transport through the molecule is assumed to be dominated by a single, spin-degenerate electronic level with energy $\varepsilon$ (measured with respect to the zero-bias Fermi energies of the leads) in the presence of one vibrational mode with frequency $\omega_0$. (A similar treatment for oscillations of the molecule's center of mass is straightforward \cite{koch}.) This is described by the Hamiltonian $H= H_\text{mol} + H_\text{leads} + H_\text{mix}$ \cite{glazman2,schoeller3,aleiner,koch}, where 
\begin{align}
H_\text{mol}= &(\varepsilon-eV_g) n_d + \frac{U}{2} n_d(n_d-1)
 \\\nonumber
&+ \lambda \hbar\omega_0 (b^\dag + b)n_d + \hbar\omega_0(b^\dag b+1/2),  \label{Hmol}
\end{align}
describes the molecular degrees of freedom,  $H_\text{leads}= \sum_{a=L,R}\sum_{\vc{p},\sigma} \epsilon_\vc{p}c^\dag_{a\vc{p}\sigma}c_{a\vc{p}\sigma}$ the leads, and $H_\text{mix}= \sum_{a=L,R}\sum_{\vc{p},\, \sigma} \left( t_a c^\dag_{a\vc{p}\sigma} d_\sigma + \text{h.c.}\right)$ the tunneling between leads and molecule. 
Here, Coulomb blockade is taken into account via the charging energy $U$. We focus on the regime of strong Coulomb blockade, $U\rightarrow\infty$, appropriate when $eV,k_BT\ll U$. The operator $d_\sigma$ ($d_\sigma^\dag$) annihilates (creates) an electron with spin projection $\sigma$ on the molecule, $n_d=\sum_\sigma d_\sigma^\dag d_\sigma$ denotes the corresponding occupation-number operator. Similarly, $c_{a\vc{p}\sigma}$ ($c_{a\vc{p}\sigma}^\dag$) annihilates (creates) an electron in lead $a$ ($a=L,R$) with momentum $\vc{p}$ and spin projection $\sigma$.

Vibrational excitations are annihilated (created) by $b$ ($b^\dag$). They couple to the electric charge on the molecule by the term $\sim n_d(b^\dag + b)$, which can be eliminated by a canonical transformation \cite{aleiner,glazman2},
leading to a renormalization of the parameters $\varepsilon$ and $U$, and of the lead-molecule coupling $t_a\rightarrow t_a \exp[-\lambda(b^\dagger-b)]$. From now on, we refer to the renormalized parameters as $\varepsilon$ and $U$. 

The coupling between molecule and leads is parameterized by the tunneling matrix elements $t_L$ and $t_R$. Here, we consider the weak coupling regime so that the energy broadening $\gamma$ of molecular levels due to $H_\text{mix}$ is small, \emph{i.e.}~$\gamma\ll k_BT,\hbar\omega_0$, and a perturbative treatment for $H_\text{mix}$ in the framework of rate equations is appropriate \footnote{In the \emph{low}-temperature regime $k_BT\ll\gamma$ a polaron instability has been predicted by Gogolin and Komnik \cite{gogolin}, and subsequently discussed controversially by Mitra \emph{et al.}~\cite{aleiner}. We emphasize that the results in the present paper concern the \emph{high}-temperature regime and are not related to the putative polaron instability.}. For simplicity, we assume a symmetric device with $t_L=t_R\equiv t_0$ and identical voltage drops of $V/2$ across each junction. (Generalizations are straightforward and do not lead to qualitatively new results.)

The molecular occupation probabilities $P^n_q$ for $n$ additional electrons  and $q$ excited phonons on the molecule  are determined from the rate equations 
\be\begin{split}\label{rateeq} 
\frac{dP^n_q}{dt}= &\sum_{n',q'}  \left[ P^{n'}_{q'} W^{n'\rightarrow n}_{q'\rightarrow q} - P^{n}_{q} W^{n\rightarrow n'}_{q\rightarrow q'} \right]\\
&-\frac{1}{\tau}\left[{\textstyle P^n_q- P^\text{eq}_q \sum_{q'} P^n_{q'}   }\right].
\end{split}\ee  
Here, the coefficient $W^{n\rightarrow n'}_{q\rightarrow q'}$ denotes the golden-rule rate for a transition $\ket{n,q}\rightarrow\ket{n',q'}$, consisting of the bare electronic tunneling rate $2\pi\rho t_0^2/\hbar$ (with $\rho$ the density of states of the leads), a lead Fermi function, possible spin factors, and a Franck-Condon (FC) matrix element, e.g.~\cite{varma,aleiner,koch}. The last term in Eq.~\eqref{rateeq}, containing the equilibrium phonon distribution $P^\text{eq}_q$, describes relaxation of the phonon excitations with relaxation time $\tau$. Within the rate-equations approach, the current can be written as $I=\sum_{n,q,q'}  P^{n}_{q}\left[ W^{n\rightarrow (n+1)}_{ q\rightarrow q';\,L}- W^{n\rightarrow (n-1)}_{ q\rightarrow q';\,L} \right]$, where the additional subindex $L$ in the rates denotes transitions occurring between the left lead and the molecule only.

\begin{figure}
  \psfrag{bias}[][][0.9]{$V\;(\hbar\omega_0/e)$}
  \psfrag{gate}[][][0.9]{$V_g\;(\hbar\omega_0/e)$}
  \psfrag{current}[][][0.9]{$I\;(2\pi e \rho t_0^2/ \hbar)$}
	\centering
		\includegraphics[width=1.0\columnwidth]{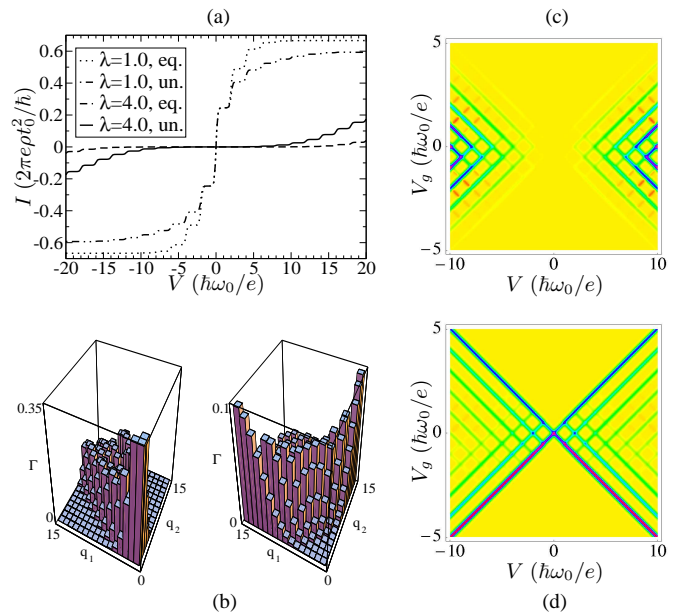}
	\caption{(Color online) (a) Current-voltage characteristics for intermediate ($\lambda=1$) and strong ($\lambda=4$) electron-phonon coupling for $\varepsilon=0$ and $k_BT=0.05\hbar\omega_0$ with (eq.) and without (un.) relaxation of phonons. We find that strong electron-phonon coupling leads to a significant current suppression at low bias voltages. This Franck-Condon blockade arises from the behavior of the Franck-Condon rates for phonon transitions $q_1\rightarrow q_2$ plotted in (b). The rates $\Gamma_{q_1q_2}$, shown for $\lambda=1$ (left) and $\lambda=4$ (right), are given in units of $2\pi \rho t_0^2/\hbar$. For strong electron-phonon coupling, transitions between low lying phonon states are exponentially suppressed. The corresponding current suppression cannot be lifted by a gate voltage, which may serve as a fingerprint of FC blockade. This is depicted in the plot of $dI/dV$ in the $V$--$V_g$ plane for unequilibrated phonons with $\lambda=4$ (c). The case of intermediate coupling with $\lambda=1$ (d) is shown for comparison. 
 \label{fig1}}	
\end{figure}

\emph{Franck-Condon blockade.}---The strong dependence of the $IV$s on the  electron-phonon coupling strength $\lambda$ is shown in Fig.~\ref{fig1}(a) for  $\lambda=1$ (intermediate coupling) and $\lambda=4$ (strong coupling), as obtained from the rate-equation approach. 
We first consider the case of $\varepsilon=0$, i.e.~the molecular single-particle level and the lead Fermi energies are aligned at zero bias. Then, for $\lambda=1$, the current increases sharply due to resonant tunneling when switching on a small bias voltage, and it exhibits the characteristic FC steps. In contrast, for $\lambda=4$ the current is \emph{significantly suppressed} at low bias voltages.

The current suppression originates from the behavior of the FC matrix elements determining the rates of phononic transitions $q_1\rightarrow q_2$. Their dependence on the electron-phonon coupling strength can be understood in terms of the overlap of two displaced harmonic oscillator wave functions. The parameter $\lambda$ determines the magnitude of this displacement in units of the oscillator length. For weak coupling, $\lambda\ll 1$, transitions mainly occur along the diagonal $q_1\rightarrow q_1$. For intermediate coupling, $\lambda\approx1$, the distribution of transition rates becomes wider, and transitions slightly off-diagonal are favored. For strong electron-phonon coupling, $\lambda \gg 1$, the distribution widens considerably and a gap of exponentially suppressed transitions between low-lying phonon states opens, see Fig.~\ref{fig1}(b). It is interesting to note that these ingredients are also crucial in the context of phonon broadening of resonant-tunneling line shapes \cite{glazman2}.

For temperatures $k_BT\ll \hbar\omega_0$, a tunneling event can increase the number of excited phonons by at most $\Delta q=\intg( 0.5e\abs{V}/\hbar\omega_0)$ due to energy conservation. Thus, for strong electron-phonon coupling and at low bias voltages, the system is trapped in a region of exponentially small transition rates. For equilibrated phonons ($\tau=0$) this suppression dominates the $IV$ until the bias voltage is high enough ($eV\sim\lambda^2\hbar\omega_0$) to escape from the blockade regime by transitions from zero phonons to highly excited phonon states. For unequilibrated phonons ($\tau=\infty$) the blockade is less rigorous, since a tunneling event can leave the molecule in an excited phonon state and  subsequent tunneling events can increase this excitation even further. 

There exist two experimental fingerprints of the FC blockade, which can be readily used to distinguish it from other low-bias current suppressions, such as off-resonance tunneling for $\varepsilon\not=0$ and Coulomb blockade. In ungated devices, the succession of FC step heights yields a fingerprint of the origin of the blockade regime. For symmetric devices with strong phonon relaxation, the presence of FC blockade is reflected in increasing step heights when leaving the voltage range of current suppression. For asymmetric devices or weak phonon relaxation, this criterion does not hold for each single step but still gives a valid tendency for the step-height succession. (We point out that systems with strong electron-phonon coupling and weak relaxation can even exhibit NDR phenomena as indicated by negative values for $dI/dV$ in Fig.~\ref{fig1}(c) in red color.) When working with gated devices, a clear distinction can be achieved by measuring $dI/dV$ as a function of both bias and gate voltage. Here, the FC blockade can be identified by an extended blockaded region in the $V$--$V_g$ plane, whereas an $\varepsilon\not=0$ suppression and Coulomb blockade can always be lifted by an appropriate adjustment of the gate voltage, see Figs.~\ref{fig1}(c),(d).

\begin{figure}
  \psfrag{bias}[][][0.9]{$V\;(\hbar\omega_0/e)$} 
  \psfrag{F}[][][0.9]{$F_\text{ex}$}
  \psfrag{l}[][][0.9]{$\lambda$} 
  \psfrag{Y}[][][0.9]{$F$}
	\centering
		\includegraphics[width=0.48\columnwidth]{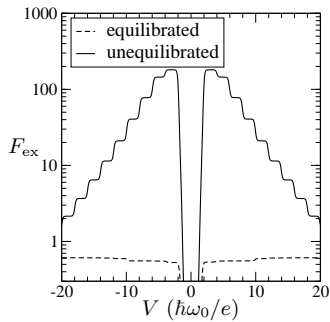}\hspace{0.2cm}
	\caption{Zero-frequency (excess) Fano factor as a function of bias voltage with $\lambda=4$ for $\varepsilon=0$ and $k_BT=0.05\hbar\omega_0$. Results are shown for equilibrated phonons ($\tau=0$)  and unequilibrated phonons ($\tau=\infty$). We find that the combined effect of strong electron-phonon coupling and weak phonon relaxation leads to giant Fano factors in the Franck-Condon blockade regime.\label{fig2}}
\end{figure}

\emph{Giant Fano factors.}---The remarkable transport properties of the system in the FC blockade regime are elucidated by computing the current fluctuations. We use Monte Carlo (MC) methods to simulate the explicit temporal dynamics of tunneling events, and rate equations  combined with a Langevin approach \cite{korotkov} to calculate the current noise
\be
S(\omega)=2\int_{-\infty}^\infty dt\, e^{i\omega t} \langle \delta I(t+t')\delta I(t') \rangle_{t'}.
\ee
Here, we restrict our discussion to symmetric junctions so that according to the Ramo-Shockley theorem \cite{blanter} the current is given by $I(t)=(I_L(t)+I_R(t))/2$. Typical results for strong and weak phonon relaxation are depicted in Fig.~\ref{fig2}, where the zero-frequency Fano factor $F_\text{ex}$ for the excess noise $S-S|_{V=0}$ is plotted as a function of bias voltage.
For equilibrated phonons, $F_\text{ex}$ is suppressed below 1 and except for weak step-like structures, it remains essentially constant in close vicinity of $F=5/9$ expected for the phononless system with one spin-degenerate level and strong Coulomb blockade \cite{nazarov}. 

In contrast, for unequilibrated phonons with $\lambda=4$ the Fano factor reaches values close to 200 as soon as the bias voltage is high enough for exciting the first phonon, and subsequently decreases stepwise at bias voltages corresponding to the opening of further phonon channels. (We note that we found even larger Fano factors of the order of $10^3$ for $\lambda=4$ and nonzero single-particle energy  $\varepsilon\not=0$.) For bias voltages sufficient to lift the FC blockade, the Fano factor returns to values of the order of 1. 

\begin{figure}
\psfrag{time}[][][0.9]{$t\;(2\pi\rho t_0^2/\hbar)^{-1}$}
\psfrag{q}[][][0.9]{$q(t)$}
\psfrag{N}[][][0.9]{$N(t)$}
	\centering
		\includegraphics[width=0.9\columnwidth]{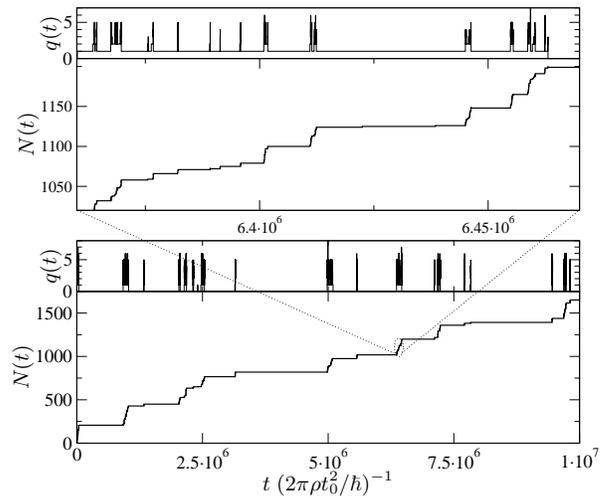}
	\caption{Number of transmitted electrons $N$ and number of excited phonons $q$ as functions of time, obtained by MC simulation for $T=0.05\hbar\omega_0$, $V=3\hbar\omega_0/e$, $\varepsilon=0$, $\tau=\infty$, and $\lambda=4$.  The upper plot shows a magnification of the marked region in the lower plot.  The results show long periods of time without current flow when the system assumes the phononic ground state, and avalanche-like current flow accompanied by phonon excitations on much shorter time scales. Self-similarity of the curves is found when magnifying the avalanche phases. \label{fig3}}
\end{figure}

The physics of the giant Fano factor originates from the combined effects of FC blockade and weak phonon  relaxation, and can be understood by considering the time-dependent dynamics of the system. Results from a MC simulation of the tunneling events for $k_B T\ll\hbar\omega_0$ and strong electron-phonon coupling are shown in Fig.~\ref{fig3} (bottom), where the number of transmitted electrons $N$ within the time interval $[0,t]$ and the number of excited phonons $q$ are plotted as functions of time $t$. 
We find that for  large periods of time the system assumes the phononic ground state $q=0$ and the current vanishes (i.e.~$N$ is constant). These phases are interrupted by \emph{avalanche-like increases} of $N$ (accompanied by phonon excitations) on much shorter time scales. When magnifying the avalanche phases, we detect self-similarity of the curves $N(t)$ and $q(t)$, see Fig.~\ref{fig3} (top): Again, one observes long times with constant $N$, now associated with $q=1$, and short avalanche phases. 

The extended phases without charge transfer are due to the suppressed transition rates for low lying phonon states in the FC blockade regime. Typically, the system spends a long time in the phononic ground state before a transition takes place. Starting from $q=0$ and for voltages $2\hbar\omega_0<e\abs{V}<4\hbar\omega_0$, where the Fano factor is maximal, energy conservation only allows the phonon transitions $0\rightarrow 0$ and $0\rightarrow 1$.  Since the rates for phonon transitions $0\rightarrow q_2$ roughly grow exponentially with $q_2$, it is favorable to increase the phonon number whenever an electronic transition occurs. Thus, as soon as the system succeeds in leaving the phononic ground state  and if phonon relaxation is weak, subsequent transitions tend to increase the phonon number even further, while the transition rates increase roughly exponentially. Then, after having left the regime of suppressed transition rates, the system can exhibit an avalanche of tunneling events, until a transition to the phononic ground state induces another long period without charge transfer, and the cycle repeats. Effectively, this leads to electron bunching, and the Fano factor can be estimated by the typical number of electrons transmitted during an avalanche, see Fig.~\ref{fig3}.

\begin{figure}
\psfrag{legendt1}[][][0.7]{$\quad eV=3.0\hbar\omega_0$}
\psfrag{legendt2}[][][0.7]{$\quad eV=6.5\hbar\omega_0$}
\psfrag{legendt3}[][][0.7]{$\quad eV=9.0\hbar\omega_0$}
\psfrag{sofomega}[][][0.9]{$S(\omega)\;(2\pi e^2 \rho t_0^2/\hbar)$}
\psfrag{softau}[][][0.9]{$S(\omega=0)\;(2\pi e^2 \rho t_0^2/\hbar)$}
\psfrag{omega}[][][0.9]{$\omega\;(2\pi\rho t_0^2/\hbar)$}
\psfrag{tau}[][][0.9]{$1/\tau\;(2\pi\rho t_0^2/\hbar)$}
	\centering
		\includegraphics[width=0.9\columnwidth]{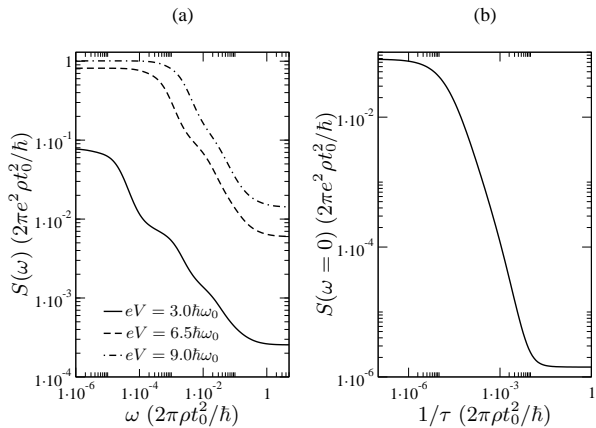}
	\caption{Noise power in the Franck-Condon blockade regime as a function of (a) frequency, and (b) phonon relaxation rate for $\varepsilon=0$, $\lambda=4$, and $k_BT=0.05\hbar\omega_0$.  In (a) the total current noise is plotted for  unequilibrated phonons. In (b) the zero-frequency noise is plotted for the bias voltage $eV=3\hbar\omega_0$.  We find (approximate) power-law behavior $S(\omega)\sim \omega^{-\alpha}$ and $S(0)\sim\tau^{1/\alpha}$ with $\alpha\approx1/2$, arising from the self-similar  avalanches over a wide range of time scales.  \label{fig4}}
\end{figure}

The remarkable self-similarity of avalanche phases due to the exponential stretching of transition rates is reflected in the relaxation-rate dependence of the Fano factor and the noise power spectrum, depicted in Fig.~\ref{fig3}. In both cases, we find approximate power-law behavior,  $S\sim\omega^{-\alpha}$ and $S\sim \tau^{1/\alpha}$ with $\alpha\approx1/2$, over many orders of magnitude in relaxation rate and frequency, respectively. (In some cases with $\lambda=4$ and $\varepsilon\not=0$, we found this power-law behavior to extend over 6 orders of magnitude in $\omega$ and $1/\tau$.)  The boundaries of the power-law scaling in the frequency domain are connected to the maximum available transition rate, and to the ``smallest'' transition  rate  describing the escape from the phononic ground state. Presumably, the weak oscillatory deviations from a pure power law arise from the discreteness of the variable $q$.

\emph{Conclusions.}---Vibrational degrees of freedom can play an important role in transport through single molecules, and lead to transport properties very different from those of conventional semiconductor or metal nanostructures. In particular, we have shown that FC physics, which is \emph{characteristic of molecules}, leads to a low-bias current suppression (FC blockade) and giant Fano factors of the order of $10^2$--$10^3$ in single-molecule devices with strong electron-phonon coupling and weak phonon relaxation. The Fano factor enhancement has been explained by avalanche-like transport of electrons.  The self-similar occurrence of  avalanches over a wide range of time scales is reflected in approximate power-law behavior of the noise as a function of frequency and relaxation rate. We finally remark that an experimental verification of this new tranport regime depends on designing a single-molecule device with strong electron-phonon coupling and weak direct vibrational relaxation. Strong electron-phonon coupling is even realized in simple diatomic molecules such as $\text{F}_2$ and $\text{Kr}_2$ for which  we estimate $\lambda\simeq4.4$ and $\lambda\simeq5.4$ \cite{herzberg}, and very likely in a plethora of larger molecules. In addition, vibrational relaxation times as large as 10ns have been observed in recent experiments \cite{LeRoy}. For these relaxation times, the regime discussed in this paper is reached for currents large compared to 10pA \footnote{This should be compared to currents in typical experiments ranging from nA to $\mu$A \cite{park}.}.
Alternatively, the effects predicted here may be relevant in artificial nano-electromechanical devices. Indeed, the  current suppression in the FC blockade may have recently been observed in such systems \cite{weig}.

\begin{acknowledgments}
We thank N. Schwentner for discussions, and C. Timm for help with the MC simulations. This work was supported in part by Sfb 290, and the Junge Akademie.
\end{acknowledgments}


\begin{thebibliography}{22}
\expandafter\ifx\csname natexlab\endcsname\relax\def\natexlab#1{#1}\fi
\expandafter\ifx\csname bibnamefont\endcsname\relax
  \def\bibnamefont#1{#1}\fi
\expandafter\ifx\csname bibfnamefont\endcsname\relax
  \def\bibfnamefont#1{#1}\fi
\expandafter\ifx\csname citenamefont\endcsname\relax
  \def\citenamefont#1{#1}\fi
\expandafter\ifx\csname url\endcsname\relax
  \def\url#1{\texttt{#1}}\fi
\expandafter\ifx\csname urlprefix\endcsname\relax\def\urlprefix{URL }\fi
\providecommand{\bibinfo}[2]{#2}
\providecommand{\eprint}[2][]{\url{#2}}

\bibitem[{\citenamefont{Schottky}(1918)}]{schottky}
\bibinfo{author}{\bibfnamefont{W.}~\bibnamefont{Schottky}},
  \bibinfo{journal}{Ann. Phys.} \textbf{\bibinfo{volume}{57}},
  \bibinfo{pages}{541} (\bibinfo{year}{1918}).

\bibitem[{\citenamefont{Blanter and B{\"u}ttiker}(2000)}]{blanter}
\bibinfo{author}{\bibfnamefont{Y.~M.} \bibnamefont{Blanter}} \bibnamefont{and}
  \bibinfo{author}{\bibfnamefont{M.}~\bibnamefont{B{\"u}ttiker}},
  \bibinfo{journal}{Phys. Rep.} \textbf{\bibinfo{volume}{336}},
  \bibinfo{pages}{1} (\bibinfo{year}{2000}).

\bibitem[{\citenamefont{Kuznetsov et~al.}(1998)\citenamefont{Kuznetsov,
  Mendez, Bruno, and Pham}}]{kuznetsov}
\bibinfo{author}{\bibfnamefont{V.~V.} \bibnamefont{Kuznetsov}} \emph{et al.},
  \bibinfo{journal}{Phys. Rev. B} \textbf{\bibinfo{volume}{58}},
  \bibinfo{pages}{R10159} (\bibinfo{year}{1998}); 
%
\bibinfo{author}{\bibfnamefont{G.}~\bibnamefont{Iannaccone}} \emph{et al.},
  \bibinfo{journal}{Phys. Rev. Lett.} \textbf{\bibinfo{volume}{80}},
  \bibinfo{pages}{1054} (\bibinfo{year}{1998}).

\bibitem[{\citenamefont{Novotny et~al.}(2004)\citenamefont{Novotny, Donarini, Flindt, and
  Jauho}}]{flindt}
  \bibinfo{author}{\bibfnamefont{T.}~\bibnamefont{Novotn\'y}} \emph{et al.},
  \bibinfo{journal}{Phys. Rev. Lett.} \textbf{\bibinfo{volume}{92}},
  \bibinfo{pages}{248302} (\bibinfo{year}{2004}).

\bibitem[{\citenamefont{Cottet et~al.}(2004)\citenamefont{Cottet, Belzig, and
  Bruder}}]{belzig}
\bibinfo{author}{\bibfnamefont{A.}~\bibnamefont{Cottet}} \emph{et al.},
  \bibinfo{journal}{Phys. Rev. B} \textbf{\bibinfo{volume}{70}},
  \bibinfo{pages}{115315} (\bibinfo{year}{2004}).

\bibitem[{\citenamefont{Park et~al.}(2000)\citenamefont{Park , Park, Lim,
  Anderson, Alivisatos, and McEuen}}]{park}
\bibinfo{author}{\bibfnamefont{H.}~\bibnamefont{Park}} \emph{et al.}, \bibinfo{journal}{Nature}
  \textbf{\bibinfo{volume}{407}}, \bibinfo{pages}{57} (\bibinfo{year}{2000});
%
\bibinfo{author}{\bibfnamefont{N.~B.} \bibnamefont{Zhitenev}} \emph{et al.},
  \bibinfo{journal}{Phys. Rev. Lett.} \textbf{\bibinfo{volume}{88}},
  \bibinfo{pages}{226801} (\bibinfo{year}{2002});
%
\bibinfo{author}{\bibfnamefont{R.~H.~M.} \bibnamefont{Smit}} \emph{et al.}, 
\bibinfo{journal}{Nature} \textbf{\bibinfo{volume}{419}},
  \bibinfo{pages}{906} (\bibinfo{year}{2002}).
%
\bibinfo{author}{\bibfnamefont{A.~N.} \bibnamefont{Pasupathy}} \emph{et al.},
  \bibinfo{journal}{cond-mat/0311150}  (\bibinfo{year}{2003}).  
%
\bibinfo{author}{\bibfnamefont{X.~H.} \bibnamefont{Qiu}} \emph{et al.}, 
\bibinfo{journal}{Phys.
  Rev. Lett.} \textbf{\bibinfo{volume}{92}}, \bibinfo{pages}{206102}
  (\bibinfo{year}{2004}).
%
\bibitem[{\citenamefont{Boese and Schoeller}(2001)}]{schoeller3}
\bibinfo{author}{\bibfnamefont{D.}~\bibnamefont{Boese}} \bibnamefont{and}
  \bibinfo{author}{\bibfnamefont{H.}~\bibnamefont{Schoeller}},
  \bibinfo{journal}{Europhys. Lett.} \textbf{\bibinfo{volume}{54}},
  \bibinfo{pages}{668} (\bibinfo{year}{2001});
%
\bibinfo{author}{\bibfnamefont{S.}~\bibnamefont{Braig}} \bibnamefont{and}
  \bibinfo{author}{\bibfnamefont{K.}~\bibnamefont{Flensberg}},
  \bibinfo{journal}{Phys. Rev. B} \textbf{\bibinfo{volume}{68}},
  \bibinfo{pages}{205324} (\bibinfo{year}{2003}).

\bibitem[{\citenamefont{Aji et~al.}(2003)\citenamefont{Aji, Moore, and
  Varma}}]{varma}
\bibinfo{author}{\bibfnamefont{V.}~\bibnamefont{Aji}} \emph{et al.},
  \bibinfo{journal}{cond-mat/0302222}  (\bibinfo{year}{2003}).

\bibitem[{\citenamefont{Mitra et~al.}(2004)\citenamefont{Mitra, Aleiner, and
  Millis}}]{aleiner}
\bibinfo{author}{\bibfnamefont{A.}~\bibnamefont{Mitra}} \emph{et al.},
  \bibinfo{journal}{Phys. Rev. B} \textbf{\bibinfo{volume}{69}},
  \bibinfo{pages}{245302} (\bibinfo{year}{2004}); \bibinfo{journal}{cond-mat/0409248} (\bibinfo{year}{2004})

\bibitem[{\citenamefont{Koch et~al.}(2004)\citenamefont{Koch, {von Oppen},
  Oreg, and Sela}}]{koch}
\bibinfo{author}{\bibfnamefont{J.}~\bibnamefont{Koch}} \emph{et al.},
  \bibinfo{journal}{Phys. Rev. B} \textbf{\bibinfo{volume}{70}},
  \bibinfo{pages}{195107} (\bibinfo{year}{2004}).

\bibitem[{\citenamefont{Glazman and Shekhter}(1988)}]{glazman2}
\bibinfo{author}{\bibfnamefont{L.~I.} \bibnamefont{Glazman}} \bibnamefont{and}
  \bibinfo{author}{\bibfnamefont{R.~I.} \bibnamefont{Shekhter}},
  \bibinfo{journal}{Sov. Phys. JETP} \textbf{\bibinfo{volume}{67}},
  \bibinfo{pages}{163} (\bibinfo{year}{1988});
%
\bibinfo{author}{\bibfnamefont{N.~S.} \bibnamefont{Wingreen}} \emph{et al.}, \bibinfo{journal}{Phys. Rev. Lett.}
  \textbf{\bibinfo{volume}{61}}, \bibinfo{pages}{1396} (\bibinfo{year}{1988}).

\bibitem[{\citenamefont{Korotkov}(1994)}]{korotkov}
\bibinfo{author}{\bibfnamefont{A.~N.} \bibnamefont{Korotkov}},
  \bibinfo{journal}{Phys. Rev. B} \textbf{\bibinfo{volume}{49}},
  \bibinfo{pages}{10381} (\bibinfo{year}{1994}).

\bibitem[{\citenamefont{Nazarov and Struben}(1996)}]{nazarov}
\bibinfo{author}{\bibfnamefont{Y.~V.} \bibnamefont{Nazarov}} \bibnamefont{and}
  \bibinfo{author}{\bibfnamefont{J.~J.~R.} \bibnamefont{Struben}},
  \bibinfo{journal}{Phys. Rev. B} \textbf{\bibinfo{volume}{53}},
  \bibinfo{pages}{15466} (\bibinfo{year}{1996}).

\bibitem[{\citenamefont{Huber and Herzberg}(1979)}]{herzberg}
\bibinfo{author}{\bibfnamefont{K.~P.} \bibnamefont{Huber}} \bibnamefont{and}
  \bibinfo{author}{\bibfnamefont{G.}~\bibnamefont{Herzberg}},
  \emph{\bibinfo{title}{Molecular Spectra and Molecular Structure}}, vol.
  \bibinfo{volume}{IV.} (\bibinfo{publisher}{Van Nostrand Reinhold Company, New
  York}, \bibinfo{year}{1979}).

\bibitem[{\citenamefont{LeRoy et~al.}(2004)\citenamefont{LeRoy, Lemay, Kong,
  and Dekker}}]{LeRoy}
\bibinfo{author}{\bibfnamefont{B.}~\bibnamefont{LeRoy}} \emph{et al.},
  \bibinfo{journal}{Nature} \textbf{\bibinfo{volume}{432}},
  \bibinfo{pages}{371} (\bibinfo{year}{2004}).

\bibitem[{\citenamefont{Weig et~al.}(2004)\citenamefont{Weig, Blick, Brandes,
  Kirschbaum, Wegscheider, Bichler, and Kotthaus}}]{weig}
\bibinfo{author}{\bibfnamefont{E. M.}~\bibnamefont{Weig}} \emph{et al.},
  \bibinfo{journal}{Phys. Rev. Lett.} \textbf{\bibinfo{volume}{92}},
  \bibinfo{pages}{046804} (\bibinfo{year}{2004}).

\bibitem[{\citenamefont{Gogolin and Komnik}(2002)}]{gogolin}
\bibinfo{author}{\bibfnamefont{A.~O.} \bibnamefont{Gogolin}} \bibnamefont{and}
  \bibinfo{author}{\bibfnamefont{A.}~\bibnamefont{Komnik}},
  \bibinfo{journal}{cond-mat/0207513} (\bibinfo{year}{2002}).


\end{thebibliography}
\end{document}